\documentclass[authoryear,preprint, review,12pt]{elsarticle}

\usepackage{amsmath}
\usepackage{caption}
\usepackage{url}

\usepackage{graphicx}
\usepackage{subcaption}
\usepackage{booktabs}
\usepackage{tabularx}
\usepackage{xcolor}
\usepackage{svg}

 \usepackage[table,xcdraw]{xcolor}
\usepackage{colortbl} 

\usepackage{multirow}
\usepackage{url}

\usepackage{longtable}

\usepackage{xltabular}

\usepackage{array}
\newcolumntype{Y}{>{\centering\arraybackslash}p{1cm}X}

\journal{IJHCS}
\begin{document}
\begin{frontmatter}

	

\title{Exploring The Interaction-Outcome Paradox: Seemingly Richer and More Self-Aware Interactions with LLMs May Not Yet Lead to Better Learning}
	\author{Rahul R. Divekar\corref{cor1}}
	\ead{divekar.rahul@gmail.com,  rdivekar@bentley.edu}
	\cortext[cor1]{Corresponding Author}

	\author{Sophia Guerra}
	\ead{sguerra@falcon.bentley.edu}

	\author{Lisette Gonzalez}
	\ead{lgonzalez@falcon.bentley.edu}

	\author{Natasha Boos}
	\ead{nboos@falcon.bentley.edu}

	\affiliation{organization={Bentley University},
	            addressline={175 Forest St.},
	            city={Waltham},
	            postcode={02452},
	            state={MA},
	            country={USA}}









\begin{abstract}
	While Large Language Models (LLMs) have transformed the user interface for learning, moving from keyword search to natural language dialogue, their impact on educational outcomes remains unclear. We present a controlled study (N=20) that directly compares the learning interaction and outcomes between LLM and search-based interfaces. We found that although LLMs elicit richer and nuanced interactions from a learner, they do not produce broadly better learning outcomes. In this paper, we explore this the ``Interaction-Outcome Paradox.'' To explain this, we discuss the concept of a cognitive shift: the locus of student effort moves from finding and synthesizing disparate sources (search) to a more self-aware identification and articulation of their knowledge gaps and strategies to bridge those gaps (LLMs). This insight provides a new lens for evaluating educational technologies, suggesting that the future of learning tools lies not in simply enriching interaction, but in designing systems that scaffold productive cognitive work by leveraging this student expressiveness.
\end{abstract}

	
	


\end{frontmatter}

\maketitle

\section{Introduction}

Higher education students' interactions with technology is constantly evolving as a result of new product invention and students' desire to keep up with new ways of learning and improving grades. However, major disruptions to these interactions occur less frequently. A past generation of popular tools like Google or Wikipedia disrupted the academic landscape by providing efficient access to relevant content \citep{duha2023chatgpt}. This brought students to the internet for learning rather than going to the library thereby prompting libraries to redefine their purpose \citep{brindley2006re}.

The landscape is now shifting from learning with search engines and online encyclopedias to learning with Large Language Models (LLMs) \citep{divekar2025can}. With over 85\% of students reporting  ChatGPT use for academic purposes \citep{Sallam2024}, this change is widely assumed to represent an  advancement in knowledge search and consumption patterns. As a result, similar to libraries in the past, search engines are redefining their core functionalities by moving from  enabling users to forage information with keywords to providing LLM-based tailored responses as the first information user interface element (e.g., Google's AI overview \citep{google2024generative}). 

However, there is limited research in examining (a) The interactional shifts that LLMs bring to the primarily search-based learning paradigm  (b) Whether those shifts lead to a better learning outcome. To fill the gap, we conduct a study where we examine whether students' interaction patterns in accessing information differs between LLM-based interfaces and search interfaces in the context of learning new topics. Further, we recognize that for most students, the goal for accessing information is to learn and produce output suitable to be submitted as assignments. Therefore, we investigate if either of these interactional experiences lead to better graded outcomes in an experimental setting.   

We present findings from a within-subjects experiment involving N=20 university students who were asked to learn two topics using a popular search and a popular LLM-based tool. They were subsequently asked to write an  essay on the topic to demonstrate their understanding. We analyze the data collected from the experiment from three angles: analysis of input while learning with each tool type, analysis of the output (essay) produced by the participants, and the analysis of the outcome (grade) that the students would receive on the essay. We use a combination of qualitative coding and automated linguistic analysis to compare differences between the interaction patterns by examining the input into each tool. For analyzing the output, we use TextEvaluator \citep{sheehan2016review, sheehan2015using}, an automated informational text difficulty analysis tool with previously identified psychometric properties and compare the results across two conditions. Finally, to analyze the outcome of the learning activity, we hired two external experts to grade the essays on a pre-determined rubric; we compare their scores across two conditions. Sections \ref{sec:methods} and \ref{sec:results} will detail the choice and suitability of the methods employed and the results in more detail. 

Overall, we discover the newly emerging interaction patterns of learning with LLMs are significantly richer indicating a shift in how cognitive resources are spent while learning online: from information synthesis (search) to writing prompts that exhibit awareness of knowledge gap and precise informational requests to fill it (LLMs). However, while the interaction patterns are significantly different, there is no difference in the outcome from a learning perspective. In other words, we present the interaction-outcome paradox where the presumed better tool (LLM) creates new ways of learning through a richer interaction but does not impact the outcome itself. We briefly discuss how students' expressive prompting can be leveraged to improve learning. 


\section{Literature Review}

\subsection{Self-Directed Learning}
A typical full-cycle learning process involves three broad activities: engaging in \textit{input}, \textit{process}, and \textit{output}. In the \textit{input} stage, a learner typically consumes information passively e.g., watching a video online. If they start to connect the information with prior knowledge, they are in the \textit{processing} stage. Finally, if they use their learning to produce new or synthesized information, they are in the \textit{output} stage. A typical university classroom  aims to expose students to all three stages. For example, a student may be introduced to a new topic in the class (input) while they are asked to think deeply about it (process) and then produce assignments (output) based on their thinking. 

Self-directed learning \citep{knowles1975self} is a process where students take ownership over the three phases and go beyond by identifying the gaps in information, finding appropriate resources, and applying learning strategies to further their expertise in a topic \citep{boyer2014self}.  A typical student who is exposed to some new knowledge in the class will imaginably use self-directed learning to explore the topic deeply to succeed in their assignments that require depth. Learning is improved when \textit{desirable difficulties} \citep{bjork2011making} such as interleaving or spacing topics apart are present in the learner's journey. In the past few decades, university students engaged in self-directed learning via search engines to find and synthesize new information that they then self-evaluated to be appropriate for the class learning outcomes. 

However, the interaction with technology and the resulting process through which students discover and interact with new external information is changing due to the rise of LLMs. In the past, students turned from libraries to search engines to find reference materials. Now, students turn to LLMs from search engines to fill knowledge gaps. In this paper, we investigate if this shift results in changes to the learning interactions and outcomes.  

\subsection{LLMs in Education}
The large-scale availability and marketing of LLM-based products such as ChatGPT, Claude, etc. have increased their popularity within the student community as a tool for a wide range of learning-related tasks such as understanding complex topics, writing, and brainstorming \citep{Albadarin2024}. Students use LLMs like ChatGPT at all levels of the Bloom's Taxonomy of Learning (a popular taxonomy that classifies learning activities into six types of action verbs \citep{krathwohl2002revision}) with primary usage of LLMs being at the \textit{create}, and \textit{understand} levels of learning \citep{divekar2025can}. 

Despite increased adoption, both students \citep{divekaraied25} and educators \citep{Kiryakova2023} see the rise in LLM usage as a threat to academic integrity as it can hamper education by allowing students to circumvent the \textit{process} of learning. Researchers have found some specific downsides of learning with LLMs: \citet{bastani2024generative} have found that specifically for math education, ChatGPT acted as a ``crutch'' that when taken away impacted students' skill. Similarly, \citet{kosmyna2025your}'s preprint has found that using ChatGPT to write essays leads to a cognitive debt with students engaging least cognitively when using ChatGPT as compared to search potentially leading to lower memory and recall.

While the negative aspects of LLM use in education have prompted some researchers to call for a re-thinking of higher education \citep{Crcek2023}, other researchers have claimed a gain in productivity and learning  as a result of LLM use \citep{dell2025cybernetic} in hybrid human-AI settings \cite{gondocs2025uncovering}. Despite positive productivity claims, a recent student survey investigating the modern technological toolkit for learning revealed that LLMs have not yet surpassed search engines as tools used for university-level education \citep{divekar2025can}. \citet{divekaraied25}  highlight the knowledge paradox as a potential reason:  students find that LLMs' inherent shortcomings like hallucinations cause students to not be able to verify LLM output due to the lack of subject-matter expertise required for verification, yet they cannot gain expertise unless they are able to verify the output. Some participants from their study mention the paradox  exceeds the benefits of a personalized learning interface with LLMs. While other students report circumventing the \textit{process} stage of learning altogether as LLMs can just give a student a tailored assignment and take them straight to the \textit{output} stage of learning.

We uniquely investigate the interactions with technology that reflect part of the learning process with each type of tool (search vs LLM) and connect it to learning outputs and outcomes in this work.

\subsection{Search and Information Retrieval for Learning}

Over the last few decades, university students have used web search engines to find new information to fulfil their educational needs. A small but directed field of HCI research focused its attention on the \textit{Search As Learning (SAL)} paradigm \citep{vakkari2016searching,hansen2016recent,ghosh2018} where researchers investigated how users interact with web search, how it varies by complexity of task, learning goals, knowledge levels etc. \citet{kuhlthau2005information}'s Information Search Process describes six stages: initiation, selection, exploration, formulation, collection, presentation, and assessment of information. Across these six stages, users switch between exploring, seeking, and documenting information to go from uncertainty to a sense of accomplishment about learning a new topic.  

However, learners and educators are now at an inflection point where information seeking and learning is being transformed from occurring primarily on web search engines to LLM-based or LLM-infused \text{answer engines} \citep{narayanan2025search} that may violate the information seeking process. Initial results such as the one discovered by \citet{narayanan2025search} show how users are discontent while learning with answer engines (or LLM-infused search) as these engines short circuit several steps of the information seeking process. For example, the researchers report that with LLMs, users miss having objective details, holistic viewpoints, control, and transparency in the process. Further, users also notice many technological shortcomings of LLM-based answer engines that, by nature, lead to incorrect information or low-quality or incorrect sources; all while making the LLM sound overconfident and overly simplistic in output. Ironically, in the case of higher education, \citet{divekaraied25} find that the need for students to provide iterative prompting, oversight, and verification when engaging with LLMs for academic work brings a sense of ownership over the final output for university students.

While SAL provides insights about interactions of learning with search engines, another subfield of investigation called Conversational Information Seeking (CIS) has directed its attention towards learning through conversational querying of technology systems. It includes theories and best-practices to engage a learner in searching and absorbing information through dialogue and question-answer based interactive systems  \citep{zamani2023conversational,dalton2022conversational, yu2025chat}. Despite best practices, most large scale and highly available conversational systems in recent years e.g., the LLM-based products like ChatGPT, are not primarily developed to encourage or support information seeking behavior \citep{vaswani2017attention}. Rather, these highly-available and well-marketed tools are often retrofitted by end-user students through iterative prompting to act as an educational technology.

While prior research has shed light on interaction patterns and educational benefit through SAL and CIS, it is unclear how the interaction patterns, outputs, and outcomes differ between the popular  new-age  LLM-based educational technology and traditional search and whether such LLM-infused advantage positively support information access \citep{shah2024envisioning}. Our study fills that gap with a comparative within subjects experiment that pits the two tools against each other connecting the input, output, and outcome characteristics compared across two tools.

\section{Methodology}
\label{sec:methods}

\subsection{Study Design}

An initial survey was sent to undergraduate and graduate student lists at the university where this research was conducted. Based on the survey, we picked N=20 students who indicated interest in the experiment and familiarity with Generative AI (GenAI) or LLM-based tools like ChatGPT. Students who indicated they were currently taking a course with the PI or anticipate taking a course in the future were excluded in the interest of fairness and validity of results. The resulting sample's demographic information is available in Table \ref{tab:demographics}. We scheduled hour-long in-person sessions for each participant. 

\begin{table}[!b] 
	\caption{Participant Demographics}
	\begin{minipage}{0.45\textwidth}
	\centering
	\begin{tabular}{|l m{2cm}|}
	\hline
	\textbf{Description} & \textbf{Number of Participants} \\ \hline
	
	Age Range & \\ 
	
	\hspace{2em} 18-20 & \hspace{1.5em}10   \\ 
	\hspace{2em} 21-23 & \hspace{2em}7 \\
	\hspace{2em} 24-26 & \hspace{2em}3 \\

	Level of Study & \\
	\hspace{2em} Undergraduate & \hspace{1.5em}13 \\
	\hspace{2em} Graduate & \hspace{2em}7 \\
	
	Gender & \\
	\hspace{2em} Female & \hspace{1.5em}11 \\
	\hspace{2em} Male & \hspace{2em}9 \\
	& \\
	
	\hline
	\end{tabular}
	\end{minipage}
	\begin{minipage}{0.45\textwidth}
	\centering
	\begin{tabular}{|l m{2cm}|}
	\hline
	\textbf{Description} & \textbf{Number of Participants} \\ \hline
	GPA (max. 4) & \\
	\hspace{2em} Above 3.5 & \hspace{1.5em}14 \\
	\hspace{2em} 3-3.5 & \hspace{2em}5 \\
	\hspace{2em} 2.5-3 & \hspace{2em}1 \\
	
	Majors or Minors & \\
	\hspace{2em} Finance & \hspace{2em}6 \\
	\hspace{2em} Management & \hspace{2em}5 \\
	\hspace{2em} Business Analytics & \hspace{2em}4 \\
	\hspace{2em} Marketing & \hspace{2em}3 \\
	\hspace{2em} Accounting & \hspace{2em}2 \\
	\hspace{2em} Other & \hspace{2em}7 \\
	
	\hline

	\end{tabular}
	\end{minipage}
	\label{tab:demographics}
	
	\end{table}

We designed a within-subjects experimental setup with randomized and counterbalanced assignments across two tool types: \textit{LLM} and \textit{Search} and two topics: \textit{How does the internet work?} and \textit{How does the power grid work?} From the initial intake survey, we found that ChatGPT and Google were the most popular LLM and Search tools. Therefore, we picked ChatGPT and Google as tools that represent LLM and Search tool types for our study. Most recent versions of these tools offer blended capabilities i.e., Google can show LLM-based search results while ChatGPT can retrieve external sources. However, versions available at the time of study did not overlap and kept the essence of an LLM and search based interaction separated. We picked topics like the \textit{internet} and \textit{power grid} with the assumption that the topics are of general interest and that our sample will have room to grasp them broadly in a short time using self-directed learning. Differences between pre- and post-essays shared later in the paper will confirm this assumption. 

\begin{figure}
    \centering
    \includegraphics[width=0.8\linewidth]{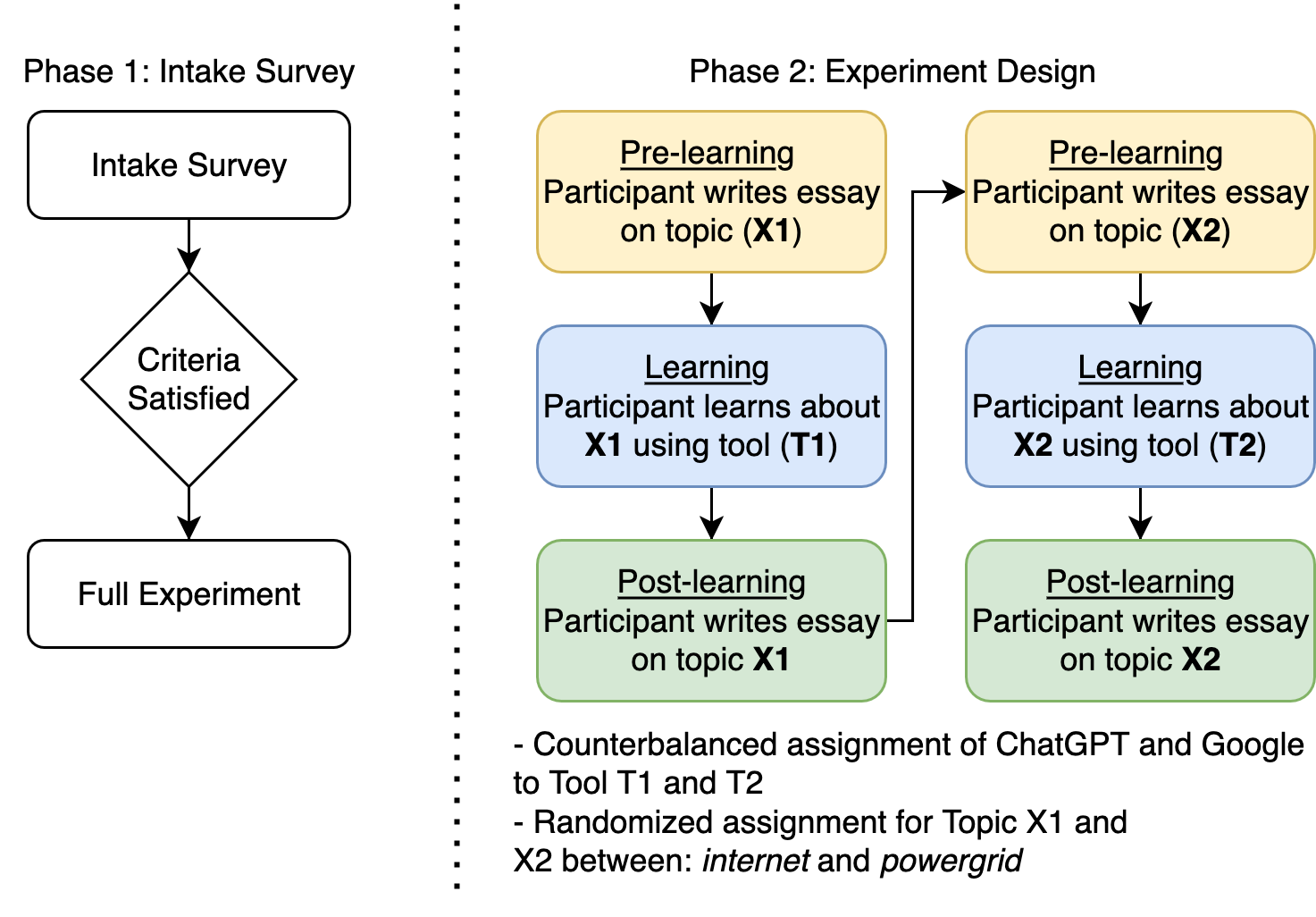}
    \caption{Study and Experiment Design}
    \label{fig:study_design}
\end{figure}

After the introductions and informed consent, all participants went through  six 10-minute time boxed activities in the order as follows. In the first activity (pre-learning), students used a simple word processer to write an essay on a given topic. The choice of topic was counterbalanced between the two mentioned above. In the second activity (learning), participants used either ChatGPT or Google (random assignment) to learn about the same topic. They were not given any directions on how to use either of the tools. In the third activity (post-learning), participants were asked to write an essay on the topic again. The topics we chose are of general interest but also involve layered concepts and technical jargon that our sample would presumably not be able to memorize in a 10-second timebox while also engaging with higher-order learning. Therefore, students were allowed to take notes in a separate file during the learning phase and refer to it during the post-learning phase with the intention that the availability of notes will reduce the need to memorize information and encourage students to engage in a higher order of learning i.e., synthesis and creation. While these 10-minute time-boxed activities and note-taking do not represent all types of in-the-wild learning or exhaustively cover the overall goals of education, it allows us to concretely measure the learning gain with the tool and also simulate a part of the self-directed learning that involves searching and synthesizing information from online sources by asking students to engage in the full cycle of learning --- \textit{input (find and read)}>\textit{process (understand)}>\textit{output (write essay)}. The three subsequent activities were mirrored with the alternative tool and topic. We concluded with a short interview.  We used screen and voice recorders to record activity. Fig. \ref{fig:study_design} shows a flow diagram summarizing the main activities. Our within-subjects pre/post design allows us to mitigate confounding factors related to individual differences in skills that we do not directly measure in this study e.g., note taking skills, search skills, and LLM prompting skills. 

The study was approved by the university IRB. Participants were compensated with a gift card.  All instruments were piloted and refined before the experiment. No data or participants from pilots were included in the final experiment.

\subsection{Data Collection and Analysis}
We analyze and present results of our data from three perspectives: Input Analysis that focuses on the extracted input features such as length, complexity, etc. to each tool; Output Analysis that focuses on the textual features of the final essays that participants wrote; and Outcome Analysis that focuses on expert-evaluated scores for the final essays. We report three significance levels of p-values: $<$ 0.001, $<$ 0.01, $<$ 0.05 and Partial Eta Squared ($\eta_p^2$) as estimate of effect size: small (0.01), medium (0.06), large (0.14).

\subsubsection{Input Analysis}
We manually reviewed each screen recording of participants' interactions across the two conditions. We transcribed the input that participants entered into the input boxes of the two interfaces and recorded the number of websites visited in the Search condition. For the purposes of this paper, input was defined as the final text (after revisions) that a user typed into the webpage form of the tool (e.g., search bar in Google and the \textit{ask anything} input area of ChatGPT), and hit enter or send. 

We derived automated metrics to establish characteristics of text input. At the simplest form, we record the number of queries typed into each tool and the average query word count of the queries. In addition, we also calculate the  average corrected type-token-ratio (CTTR) per participant using \citet{lex}'s implementation, a common linguistic metric used to quantitatively compare the variety of vocabulary used within the two text formations. The CTTR compares the total number of different of words (types) to the total number of words in any text. A higher CTTR therefore indicates greater, richer, or more diverse vocabulary observed in the text. We used the corrected form of TTR as our initial analysis showed a difference in token counts across two conditions.

We conducted a Repeated Measures MANOVA to test for overall effects of the two conditions on three dependent variables (number of inputs, average CTTR, and average word count) for 20 participants; additional univariate analysis was conducted if MANOVA showed statistically significant results.

To gain qualitative insight into user inputs, we developed a set of nine categories through an iterative process reflecting the distinct affordances of each tool. As shown in Table \ref{tab:qual-codes}, a single input could be assigned multiple categories. To quantify the prevalence of these behaviors across conditions, we calculated a Dominance Score for each category (see Eq. \ref{eq:dominance}). This score ranges from -1 (a behavior exclusive to Search) to 1 (exclusive to LLM), allowing us to systematically identify the most distinctive interaction patterns between the two interfaces. Given the difference in raw counts between LLM and Search codes, we also compute a Normalized Dominance Score. In this version, instead of using raw counts as in Equation \ref{eq:dominance}, we use normalized counts. Normalization for each condition is defined in Equation \ref{eq:normalization}

 \begin{equation}
	\label{eq:dominance}
	\text{Dominance Score} = \frac{\text{LLM Count} - \text{Search Count}}{\text{LLM Count} + \text{Search Count}}
	\end{equation}

	\begin{equation}
		\label{eq:normalization}
		\text{Normalized count} = \frac{\text{Count of code in a condition}} {\text{Total Count for that condition}}
		\end{equation}

Our experiment protocol allowed a maximum of 10 minutes for each section including to search and to write the final essay; however, many did not take the entire time. We recorded the time taken to complete each step i.e., to search (learn) and write about the topic post learning. We conducted a Repeated Measures MANOVA to find differences across two conditions; additional univariate analysis was conducted if MANOVA showed statistically significant results.


\subsubsection{Output Analysis}
We analyze the text that was produced by students in the post learning activity using automated tools to compute textual characteristics and overlap with existing sources on the web.

We used TextEvaluator \citep{sheehan2016review, sheehan2015using}, an engine with previously evaluated psychometric properties that,  given an informational  text, calculates its text complexity score on a scale of 100 to 2000 based on linguistic features such as: sentence structure, vocabulary difficulty, connections across ideas, and organization. Each of these features have subcomponents. For example, connections across ideas is determined by lexical cohesion, interactive/conversational style, and level of argumentation. Each of these subcomponents impact the score e.g., a higher level of argumentation increases complexity of the text \citep{napolitano2015online}. In addition, it also reports basic textual features such as total number of words, sentences, and paragraphs along with average words per sentence and average words per paragraph. Therefore, it gives a holistic picture of the textual characteristics of informational text. We use it to identify if the learning tool (LLM vs Search) has an effect on the final essay that the students produce by comparing the text complexity and other basic textual features such as the number of words, sentences, and paragraphs to determine the appearance of the essays. We ran each essay through TextEvaluator and gathered the metrics. Then, we used repeated measures MANOVA to evaluate differences in metrics across two conditions; additional univariate analysis was conducted if MANOVA showed statistically significant results.

While there are several plagiarism detection tools available, we test for textual overlap with existing sources using Grammarly's plagiarism tool. Given its prevalence and availability, it satisfies our goal to approximate the amount of overlap that practitioners will find when using off-the-shelf tools. We use repeated measures ANOVA to find differences in plagiarism of final output dependent on the condition. Since students could have their notes open, we wanted to make sure that most essays were not directly copy-pasted from the LLM/Search results into notes and then into the final document thereby affecting the overlap. To resolve that, we manually classified each output writing session into one of the three buckets: minimally, somewhat, or mostly copied from notes and report it in the results section.





\subsubsection{Outcome Analysis}
We employed two independent expert human raters to score the text across seven dimensions: accuracy of information, depth of information, breadth of information, relevancy of information, flow and organization, spelling, grammar, and vocabulary, and demonstration of topic expertise. Raters could choose a point between 5 (max) and 1 (min) to rate the essay as per rubric shown in Table \ref{tab:rubric} where 5 meant better. We picked these dimensions for two reasons. One, they can broadly be applied to college-level writing on technical subjects such as the topics used in our study. Two, previous research has criticized LLMs and Search to produce inaccurate,  shallow, or irrelevant content \citep{narayanan2025search}. Rubrics related to depth, breadth, expertise, and relevancy of information aim measure if this effect transpires into the final essay. Further, LLMs provide answers in full paragraphs as compared to search engines potentially relieving a student from having to synthesize information from distinct sources as in the case of search potentially affecting flow, organization, grammar, and vocabulary used in the final essay which are also assessed by raters in this study to identify any transpired tool effects.  

\begin{xltabular}{\textwidth}{p{2.1cm}p{0.8cm}X}
	\caption{Rubric Provided to Independent Raters to Evaluate Student Essays}
	\label{tab:rubric} \\
	\textbf{Metric} & \textbf{Score} & \textbf{Description} \\
	\hline
	\multirow{5}{2cm}{Accuracy of Information} & 5 & All of the points in the essay are accurate  \\
								& 4 & Most of the points in the essay are accurate  \\
								& 3 & Equal mix of accurate and inaccurate information \\
								& 2 & Most of the points in the essay are inaccurate  \\
								& 1 & All of the points in the essay are inaccurate \\ \hline
	
	\multirow{5}{2cm}{Depth of Information} & 5 & The essay addresses topic in sufficient depth covering each facet of technology in detail \\
							& 4 & The essay addresses topic in somewhat sufficient depth \\
							& 3 & There is some depth of information to the topic \\
							& 2 & There is little depth of information to the topic \\
							& 1 & There is no depth of information to the topic \\ \hline							
	\multirow{5}{2cm}{Breadth of Information} & 5 & The essay addresses topic in sufficient breadth covering many facets of the technology \\
							& 4 & The essay addresses topic in somewhat sufficient breadth \\
							& 3 & There is some breadth of information to the topic \\
							& 2 & There is little breadth of information to the topic \\
							& 1 & There is no breadth of information to the topic \\ \hline
	\multirow{5}{2cm}{Relevancy of Information}& 5 & All of the information is relevant to the main topic \\
							& 4 & Most of the information is relevant to the main topic \\
							& 3 & Some of the information is relevant to the main topic \\
							& 2 & Little information is relevant to the main topic \\
							& 1 & No information is relevant to the main topic \\ \hline
	\multirow{5}{2cm}{Flow and Organization} & 5 & Excellent flow and organization. Clear introduction, logical progression of ideas, effective conclusion. Smooth transition between paragraphs  \\
						& 4 & Good flow and organization. Mostly clear introduction, logical progression, and conclusion. Adequate transitions \\
						& 3 & Fair flow and organization. Introduction, progression, and conclusion are unclear or disjointed in parts. Weak transitions. \\
						& 2 & Poor flow and organization. Lacks clear introduction, progression, or conclusion. Transitions are missing or ineffective \\
						& 1 & The essay's organization and flow is detrimental to its reading and grading \\ \hline
	\multirow{5}{2cm}{Spelling, Grammar, Vocabulary} & 5 & Excellent mechanics. No errors in spelling, grammar, punctuation, or formatting.  \\
						& 4 & Good mechanics. Few errors that do not significantly impede readability.  \\
						& 3 & Fair mechanics. Errors are frequent enough to somewhat impede readability.  \\
						& 2 & Poor mechanics. Errors are severe and frequent enough to significantly impede readability. \\
						& 1 & The essay's spelling and grammar is detrimental to its reading and grading \\ \hline
	\multirow{5}{2cm}{Expertise} & 5 & Authors demonstrate excellent understanding and expertise of the topic  \\
						& 4 & Authors demonstrate good understanding and expertise of the topic  \\
						& 3 & Authors demonstrate fair understanding and expertise of the topic  \\
						& 2 & Authors demonstrate poor understanding and expertise of the topic  \\
						& 1 & Authors demonstrate no understanding or expertise of the topic \\ \hline

\end{xltabular}

The raters are not part of this research group and were chosen given their qualifications of being employed as a science teacher at a local school. The raters were compensated for their work. The raters were provided with the context of the study and that the rating was for research purposes will not affect students' academic performance to prevent skewed scores. Raters were given separate files for each pre and post essay that students wrote across the two conditions (N=80 files). Any metadata such as filenames were scrubbed or randomized to hide  identifiers of the condition to which the essay belonged to reduce bias. The raters first independently scored each file across the rubric and then used a consensus-through-discussion based scoring method to resolve discrepancies between scores. Scores were left unchanged if consensus was not reached. We use average of two raters as final score for each dimension. The total score for an essay is a sum of scores received across all 7 dimensions normalized by dividing it by 7. We  use Repeated Measures MANOVA to determine effects of the two tools on seven grading dimensions with post-hoc univariate analysis to interpret MANOVA results.  

	%

%
%

\section{Results}
\label{sec:results}
\subsection{Input Analysis}
\begin{figure}
	\centering
\begin{subfigure}{0.4\linewidth}
    \includegraphics[width=\linewidth, trim=0 0 0 0.8cm, clip]{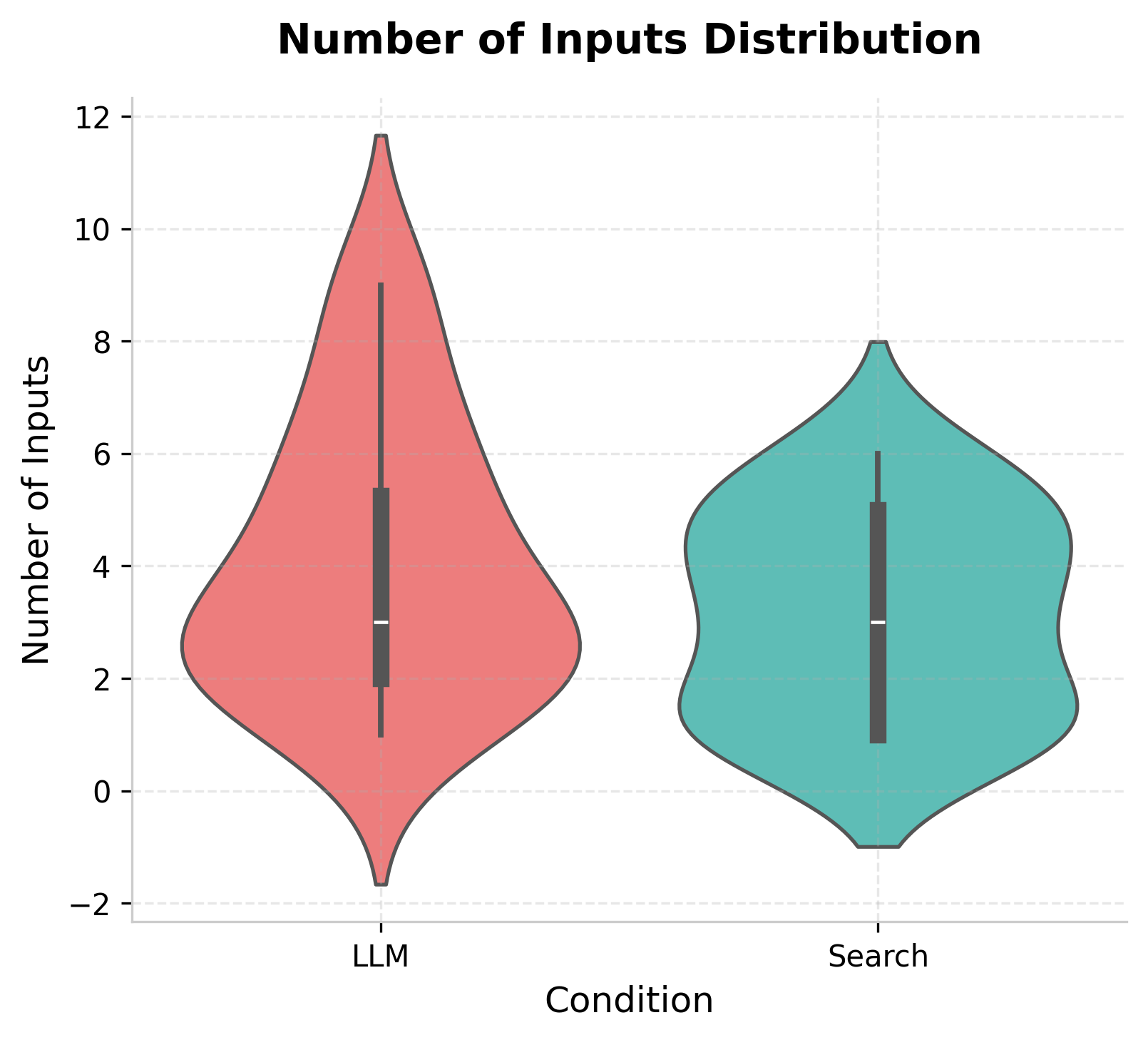}
    \caption{Number of Inputs}
    \label{fig:input_count}
\end{subfigure}

\begin{subfigure}{0.4\linewidth}
    \includegraphics[width=\linewidth, trim=0 0 0 0.8cm, clip]{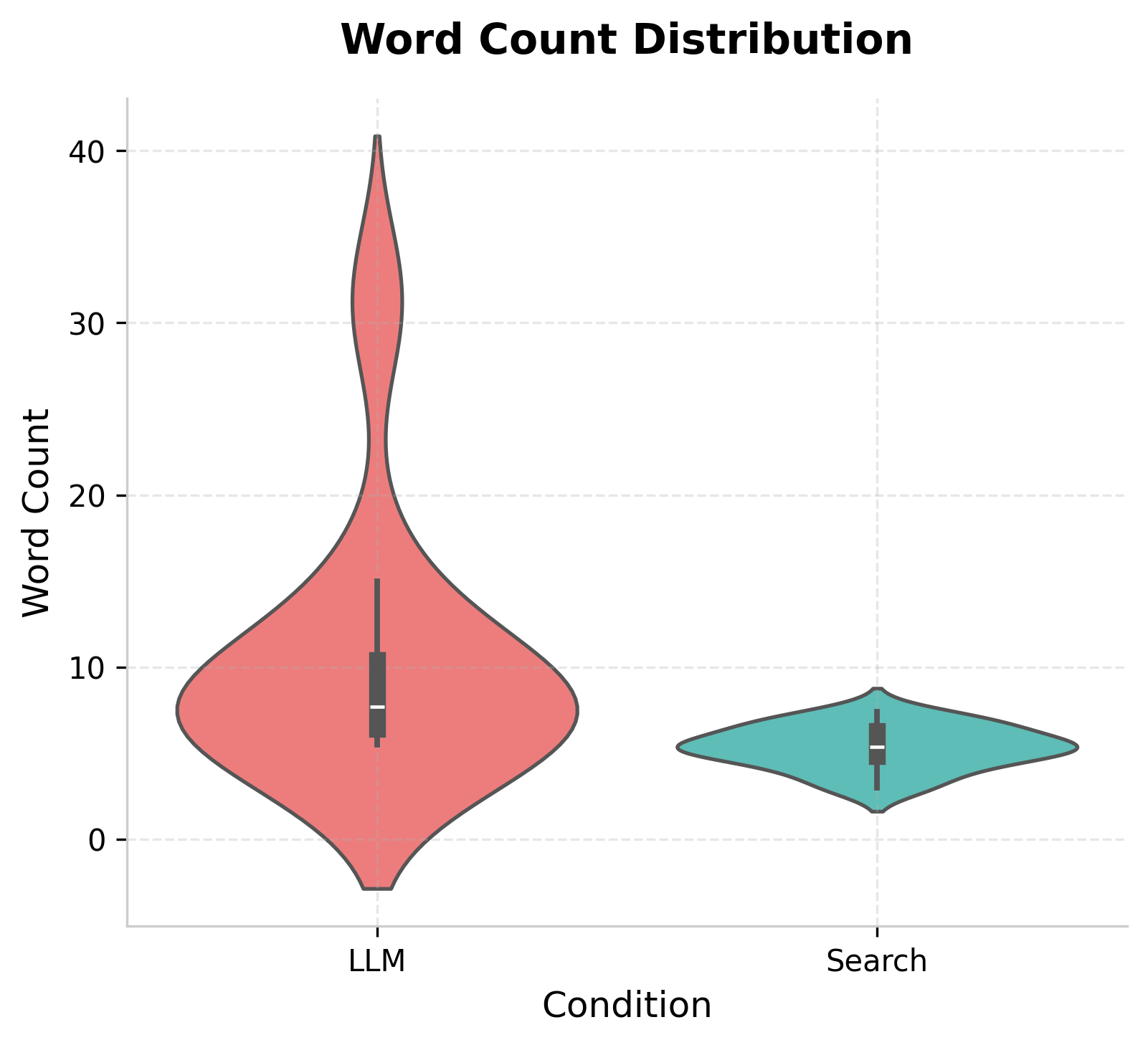}
    \caption{Word Count Per Input}
    \label{fig:word_count}
\end{subfigure}

\begin{subfigure}{0.4\linewidth}
    \includegraphics[width=\linewidth, trim=0 0 0 0.8cm, clip]{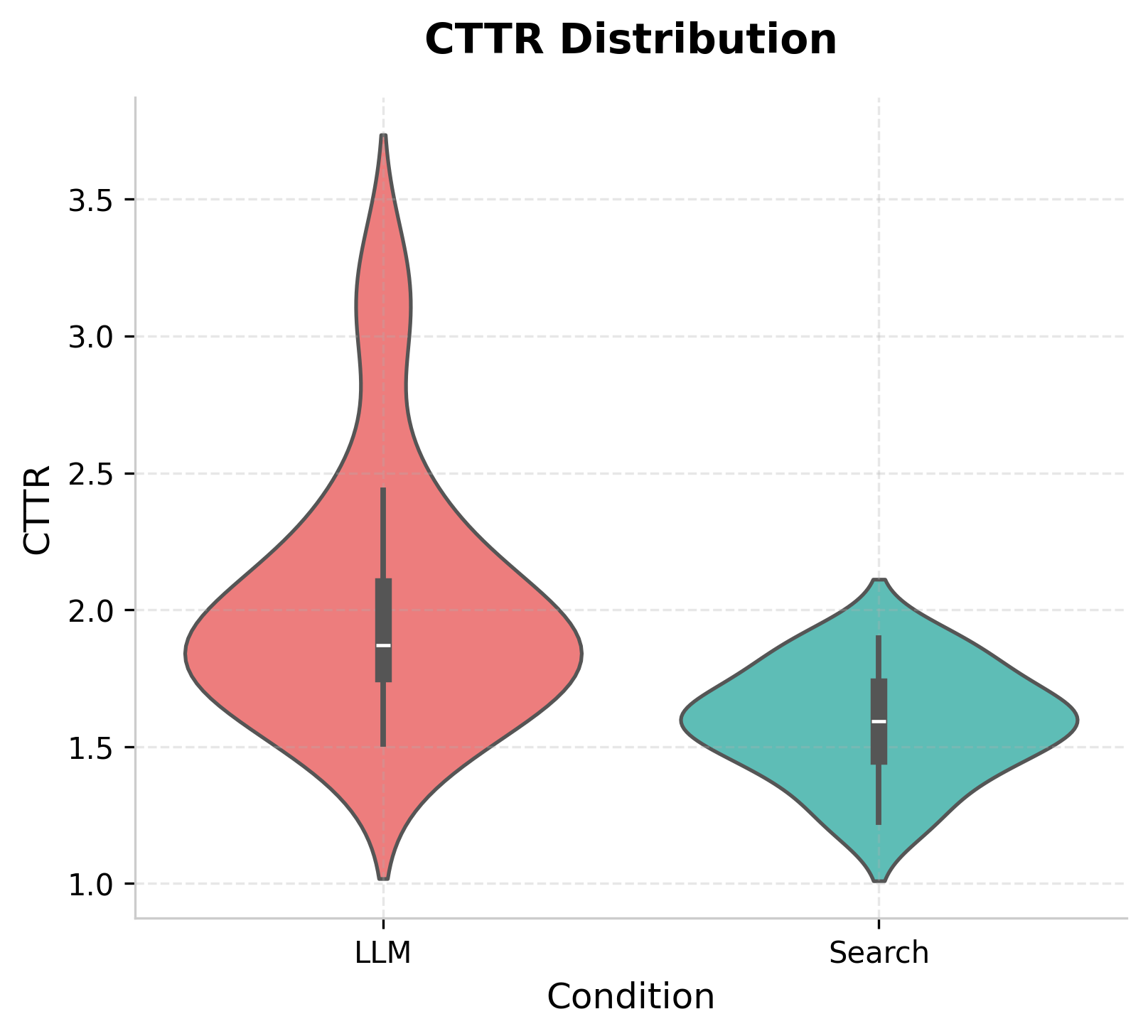}
    \caption{Corrected Type-Token Ratio of Input}
    \label{fig:cttr}
\end{subfigure}
\caption{Comparing Input Characteristics Across Two Conditions}
\label{fig:inputfeatures}
\end{figure}

Table \ref{tab:input-analysis} shows statistically significant combined effect of the two tools on Word Count, Corrected Type Token Ratio (CTTR), and the Number of Inputs with a large effect size. Further univariate analysis in the same table shows that Word Count and Corrected Type Token Ratio (CTTR) had a statistically significant difference between the two conditions with a large effect size for Word Count and CTTR but a trending non-statistically significant difference on Number of Inputs. Fig. \ref{fig:inputfeatures} shows the distributions with the LLM conditions having a higher and more distributed Word Count (LLM: 10.48$\pm$7.6, Search: 5.34$\pm$1.25), CTTR (LLM: 2.02$\pm$0.45, Search:1.59$\pm$0.2) and Number of Inputs (LLM: 4.1$\pm$2.4, Search:3.15$\pm$1.8). 

Repeated Measures MANOVA found no statistically significant difference across the two conditions on the time spent with the tool to search information or the time taken to write the final essay: Pillai's Trace = .095,F(2,18)=.944, p=0.408.  However, we note that in the search condition, participants visited an average of 5.45$\pm$2.82 websites for their learning session.

Table \ref{tab:qual-codes} summarizes the distribution of query types across LLM and Search methods, highlighting notable differences in how users interact with each system. We note that students engage in a different variety of input types with LLMs as compared to search engines with three categories (output control, reference to previous text, and context) being unused in the search condition and two category (multimedia request, keywords) being unused in the LLM condition. 

\begin{table}[]
	\caption{Repeated Measures MANOVA and Univariate Analysis of Effect of Condition \\ on Input Characteristics (***p \textless .001, **p\textless{}0.01, *p\textless{}0.05)}
	\begin{tabularx}{\linewidth}{p{3cm}p{2cm}XXXX}
		\textbf{Test}              & \textbf{Dependent Variable} & \textbf{F} & \textbf{df} & \textbf{p}         & \textbf{$\eta_p^2$} \\ \toprule
	\textbf{Multivariate Test} &                             &            &             &                    &                              \\
	Pillai's Trace             & Combined                  & 9.521      & 3,17        & \textless{}0.001\footnotesize{***} & .627                         \\ 
	Wilks' Lambda              & Combined                  & 9.521      & 3,17        & \textless{}0.001\footnotesize{***} & .627                         \\ \midrule
	\textbf{Univariate Tests}  &                             &            &             &                    &                              \\ 
							   & Word Count                  & 8.871      & 1,19           & .008\footnotesize{**}              & .318                         \\
							   & CTTR                        & 16.967     & 1,19           & \textless{}0.001\footnotesize{***} & .472                         \\
							   & Number of Inputs            & 3.944      & 1,19           & 0.062              & .172          \\  
	\bottomrule
	\end{tabularx}
	\label{tab:input-analysis}
	\end{table}

\begin{table}[]
	\centering
	\caption{Categorization of Input Query Text and Across LLM and Search Tools. Dominance score shows exclusivity of usage on a scale from LLM only (1) to Search only (-1) conditions calculated using Eq. \ref{eq:dominance}, \ref{eq:normalization}}
	\renewcommand{\arraystretch}{1.5}
	\begin{tabularx}{\linewidth}{p{1.7cm}Xp{0.7cm}p{0.9cm}p{1.1cm}p{1.2cm}}
		\toprule
	Category                          & Description                                                                                                                                             & LLM Count & Search Count & Domina- ance Score& Normal- ized Dominance Score $\downarrow$\\ \midrule

	Output Control               & Text attempts to control output. E.g., ``Make sure you start with...'', ``elaborate''                                                                       & 12        & 0            & 1    &    1  \\

	Reference to Previous Text & Text attempts to refer to previous output. E.g., ``can you take all the information you just gave me and put it in one flowing, cohesive essay together'' & 11        & 0            & 1   & 1      \\

	Context                       & Text provides more information about the larger goal. E.g., ``This is my essay topic...'', ``... your audience does not know anything about the topic''            & 2         & 0            & 1      & 1    \\

	Difficulty Control           & Text attempts to control difficulty of output. E.g. ``In the most simplest terms..''                                                                     & 15        & 7            & 0.36   & 0.06       \\

	Request Example              & Text requests an example. E.g., ``explain with an example how...''                                                                                        & 2         & 1            & 0.33   & 0.02       \\

	How/why Question            & Text contains a how or why question. E.g., ``how does the internet work''                                                                                 & 41        & 25           & 0.24    & -0.08     \\

	What question                & Text contains a what question. E.g., ``what is the internet''                                                                                             & 30        & 21           & 0.18        & -0.15  \\

	Keywords                      & Text contains only keywords rather than full sentences. E.g., ``wind city''                                                                               & 0         & 4            & -1   & -1       \\

	Multi- media Request           & Text requests multimedia. E.g.,``... educational video''                                                                                                  & 0         & 1            & -1  & -1    \\
	\bottomrule   
	\end{tabularx}
	\label{tab:qual-codes}
	\end{table}

\subsection{Output Analysis}

Using metrics derived from TextEvaluator, repeated measures MANOVA showed no significant difference in the final essays produced in either conditions across dependent measures: text complexity, word count, sentence count, and paragraph count.  Pillai’s Trace = .236, F(4,16)=1.237, p=0.335.

\begin{figure}
    \centering
    \includegraphics[width=0.5\linewidth, trim=0 0 0 0.8cm, clip]{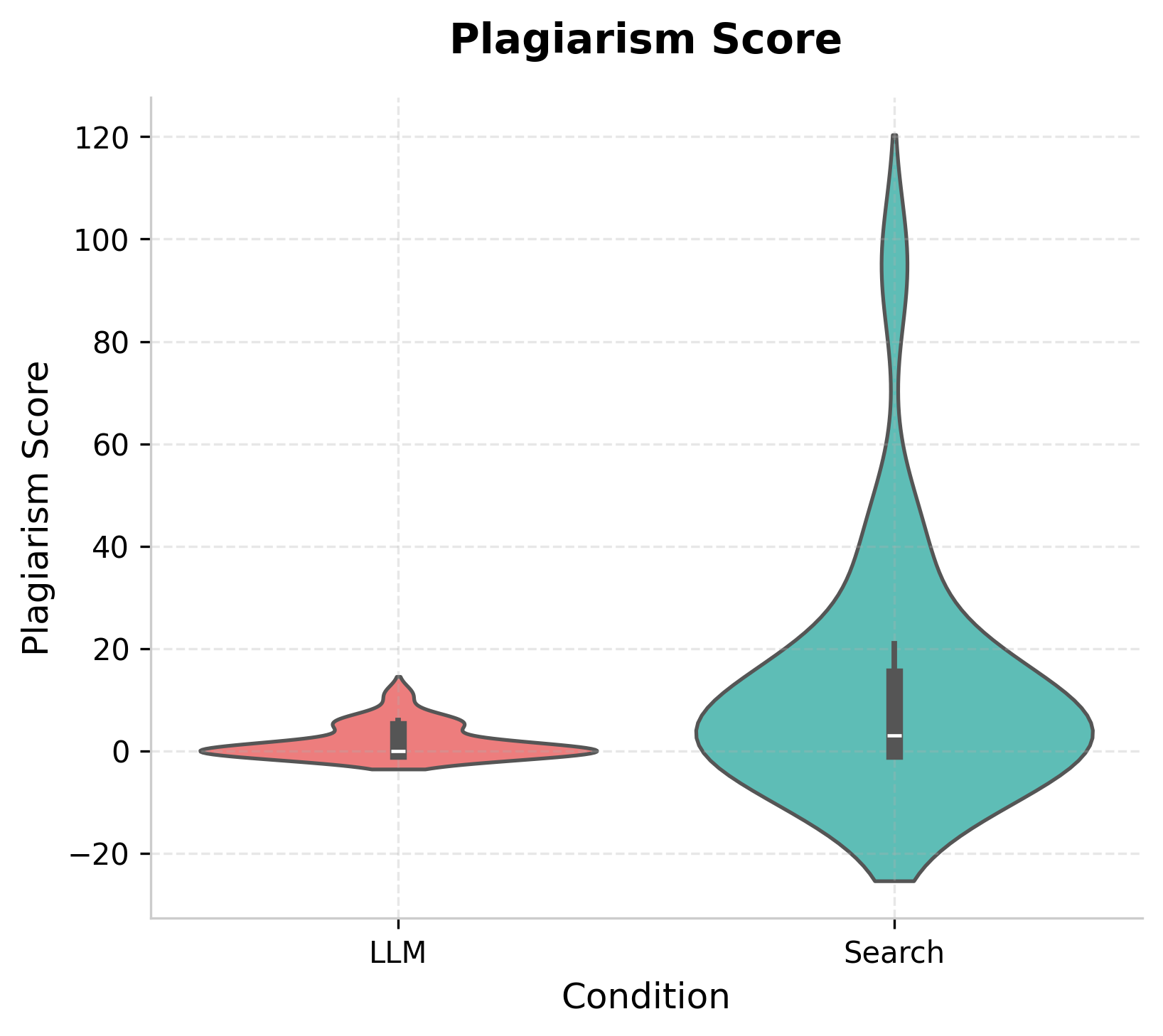}
    \caption{Plagiarism Scores of Final Output across Two Conditions}
    \label{fig:plagiarism_violin}
\end{figure}

Manual analysis found that across both conditions only 4 essays were mostly copied from notes (3 in LLM condition, 1 in Search) indicating that when students consulted their notes to write the final output, most did not outright copy-paste from the notes. Repeated measures ANVOA found a statistically significant difference in plagiarism percentages of the final essay across the search and LLM conditions with a medium to large effect size: F(1,19) = 5.528, p=0.033, partial eta squared=.217. Fig. \ref{fig:plagiarism_violin} shows the two distributions with search condition having a higher mean than plagiarism condition however with a large standard deviation (12.65$\pm$23.10) versus LLM condition (1.9$\pm$3.23).

\subsection{Outcome Analysis}

\begin{figure}
		\centering
		\includegraphics[trim=0 1.5cm 0 0, clip, width=\linewidth]{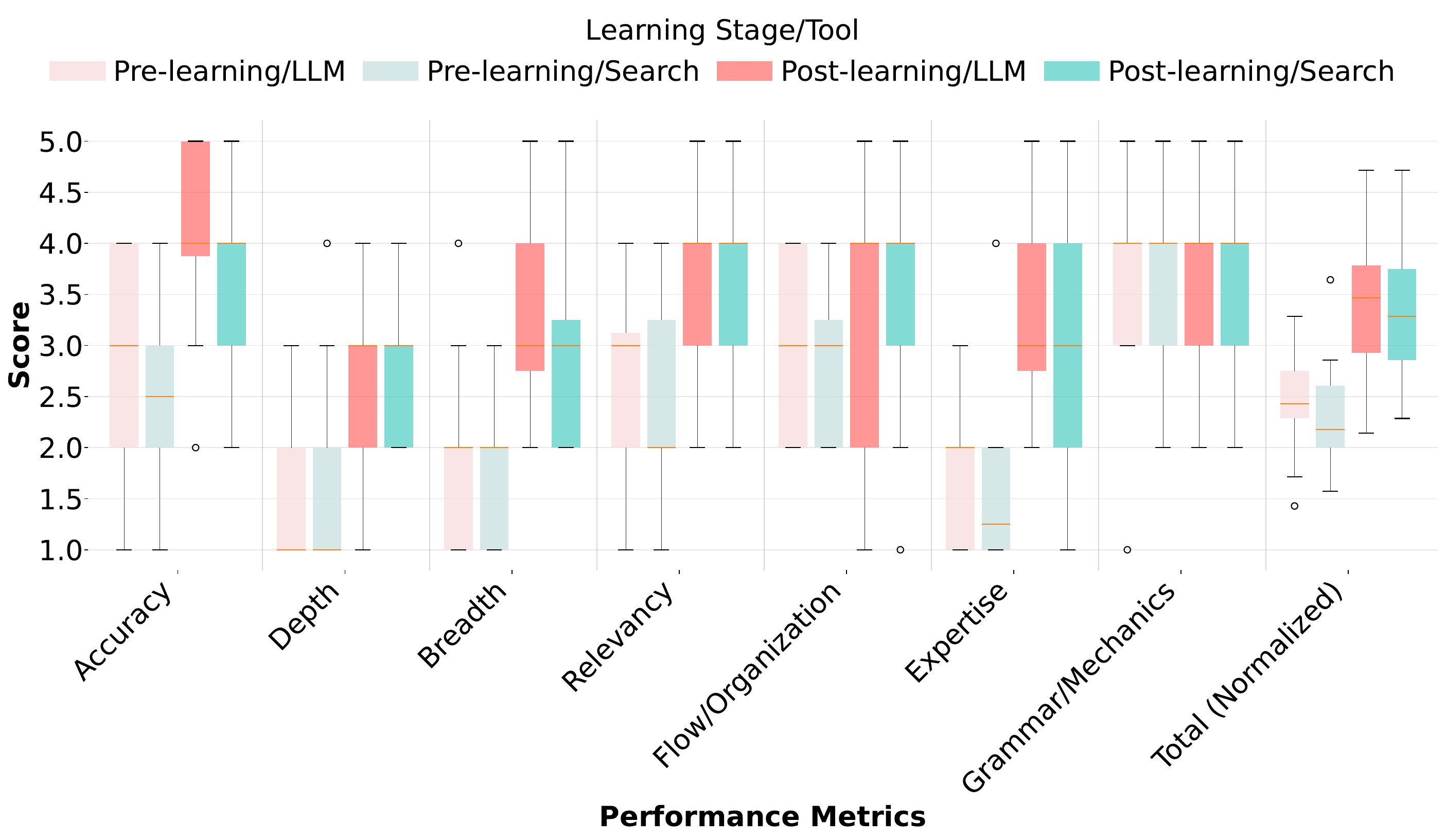}
		\caption{Scores Across Seven Dimensions Compared By Condition and Learning Stage}
	
		\label{fig:scores_box}
	\end{figure}

 Fig. \ref{fig:scores_box} shows the average scores of two raters as the final score for essays plotted across four stages: pre-LLM, post-LLM, pre-Search, and post-Search.

Table \ref{tab:pre-post-search} and Table \ref{tab:pre-post-llm} are results from Repeated Measures MANOVA and subsequent univariate analysis  that show a statistically significant increase  between pre and post scores for both tools across all dimensions except \textit{Flow and Organization} and \textit{Spelling, Grammar, and Mechanics}. However, as seen in Table \ref{tab:post-post-llm-search}, the post scores compared between two tools do not significantly differ from each other. Additional univariate analysis showed only a statistical significance in Accuracy (LLM: 4.08$\pm$0.86, Search: 3.6$\pm$0.99) with large effect.


	\begin{table}[]
		\caption{Repeated Measures MANOVA and Univariate Analysis of \\ Effect of Pre vs Post Scores in Search Condition (***p \textless .001, **p\textless{}0.01, *p\textless{}0.05)}
		\begin{tabularx}{\linewidth}{p{3cm}p{2cm}XXXX}
			\textbf{Test}              & \textbf{Dependent Variable} & \textbf{F} & \textbf{df} & \textbf{p}         & \textbf{$\eta_p^2$} \\ \toprule
		\textbf{Multivariate Test} &                             &            &             &                    &                              \\
		Pillai's Trace             & Combined                  & 6.847      & 7,13        & 0.002\footnotesize{**} & .787                         \\ 
		Wilks' Lambda              & Combined                  & 6.847      & 7,13        & 0.002\footnotesize{**} & .787                         \\ \midrule
		\textbf{Univariate Tests}  &                             &            &             &                    &                              \\ 
			& Accuracy & 12.677      & 1,19           & 0.002\footnotesize{**}              & .400                         \\
			& Depth & 53.441     & 1,19           & \textless{}0.001\footnotesize{***} & .738                         \\
			& Breadth & 22.773      & 1,19           & \textless{}0.001\footnotesize{***}              & .545          \\  
			& Relevancy & 27.092      & 1,19           & \textless{}0.001\footnotesize{***}              & .588          \\  
			& Flow and Organization & 4.717      & 1,19           & 0.043              & .199          \\  
			& Expertise & 30.921      & 1,19           & \textless{}0.001\footnotesize{***}              & .619          \\  
			& Spelling, Grammar, Mechanics & 0      & 1,19           & 1              & 0   \\  
			& Total Score & 33.421      & 1,19           & \textless{}0.001\footnotesize{***}              & .638   \\  
		\bottomrule
		\end{tabularx}
		\label{tab:pre-post-search}
		\end{table}

		\begin{table}[]
			\caption{Repeated Measures MANOVA and Univariate Analysis of Effect \\ of Pre vs Post Scores in LLM Condition (***p \textless .001, **p\textless{}0.01, *p\textless{}0.05)}
			\begin{tabularx}{\linewidth}{p{3cm}p{2cm}XXXX}
				\textbf{Test}              & \textbf{Dependent Variable} & \textbf{F} & \textbf{df} & \textbf{p}         & \textbf{$\eta_p^2$} \\ \toprule
			\textbf{Multivariate Test} &                             &            &             &                    &                              \\
			Pillai's Trace             & Combined                  & 5.167      & 7,13        & 0.005\footnotesize{**} & .736                         \\ 
			Wilks' Lambda              & Combined                  & 5.167      & 7,13        & 0.005\footnotesize{**} & .736                         \\ \midrule
			\textbf{Univariate Tests}  &                             &            &             &                    &                              \\ 
				& Accuracy & 18.986      & 1,19           & \textless{}0.001\footnotesize{***}              & .500                         \\
				& Depth & 24.455     & 1,19           & \textless{}0.001\footnotesize{***} & .563                         \\
				& Breadth & 19.464      & 1,19           & \textless{}0.001\footnotesize{***}              & .506          \\  
				& Relevancy & 14.658      & 1,19           & 0.001\footnotesize{**}              & .435          \\  
				& Flow and Organization & .629      & 1,19           & .437              & .032          \\  
				& Expertise & 24.511      & 1,19           & \textless{}0.001\footnotesize{***}              & .563          \\  
				& Spelling, Grammar, Mechanics & .322      & 1,19           & 0.577              & .017   \\  
				& Total Score & 23.877      & 1,19           & \textless{}0.001\footnotesize{***}              & .557   \\  
			\bottomrule
			\end{tabularx}
			\label{tab:pre-post-llm}
			\end{table}

\begin{table}[]
	\caption{Repeated Measures MANOVA Post vs Post Score Comparison Across Two Conditions (***p \textless .001, **p\textless{}0.01, *p\textless{}0.05)}
	\begin{tabularx}{\linewidth}{p{2.8cm}p{2cm}XXXX}
		\textbf{Test}              & \textbf{Dependent Variable} & \textbf{F} & \textbf{df} & \textbf{p}         & \textbf{$\eta_p^2$}	 \\ \toprule
	\textbf{Multivariate Analysis} & & & & & \\
	Pillai's Trace             & Combined                  & 2.520      & 7,13        & 0.071 & .576                         \\ 
	Wilks' Lambda              & Combined                  & 2.520      & 7,13        & 0.071 & .576                         \\ \midrule
	\textbf{Univariate Analysis} & & & & & \\
	& Accuracy  & 4.834& 1,19 & .040\footnotesize{*}& .203 \\
	& Depth  & .603& 1,19 & .447 & .031\\
	& Breadth  & .472& 1,19 & .500& .024\\
	& Relevancy  & .00& 1,19 & 1.00& .000\\
	& Flow and Organization  & .672& 1,19 &.422 & .034 \\
	& Expertise  &.031 & 1,19 & .863& .002\\
	& Spelling, Grammar, and Mechanics  &.699 & 1,19&.413 & .036\\
	& Total  & .080 & 1,19& .781& .004\\

	\bottomrule
	\end{tabularx}
	
	\label{tab:post-post-llm-search}
	\end{table}

\section{Discussion, Implications, and Next Directions}
\label{sec:discussion}

\subsection{LLMs elicit more diverse, richer input characteristics than search}

A defining feature for a novel interface are the input characteristics and the affordances it creates to elicit different types of input. While both web search and LLMs were primarily used in text-based unimodal mode, the type of affordances that they provide elicit different input characteristics from users. In our study we saw that users interacting with LLMs enter text that has not only more words but also more types of words showing lexical richness their in their input, while the input to search contains fewer words and less lexical richness indicating comparative superficiality in input. 

Both tools elicit a different variety of inputs. For example, in the search condition, users  limited themselves to mostly asking how, why, and what questions. Whereas, for LLMs users ask to personalize the generated output using output control terms such as ``make sure you start with...'' or request examples or elaboration of a concept they are trying to learn. This is possible because the affordance of a continuous interface for learning where all individual inputs and outputs are smoothly woven into the same interaction and LLM's ability to generate text on-demand. Since we did not provide any training to our sample, we assume that users recognize these affordances provided by LLMs, or they are prevalent within the sample as prior research has shown students primarily learn how to use LLMs by trial and error and often without any formal training \citep{divekar2025can} pointing again to the affordance of the interface being the enabling factor for richer input characteristics. Students leverage these affordances to improve learning by asking the LLM to manipulate and personalize output. Students leverage LLM's ability of remembering conversational history as evident by the exclusive use of inputs that are of the type \textit{context, reference to previous text, and difficulty control} that attempt to manipulate previous output from LLM rather than only expose themselves to new information.

In contrast, search engines provide links that lead to third-party websites authored independently of a user's learning journey i.e., they do not connect to a user's prior reading or cognitive state. In contrast to the single interface of the LLM, in the search condition, students had to access large number of independently authored webpages (5.45$\pm$2.82) to get similar information. Here, from a student interaction perspective, any follow-up search becomes a new session as it lacks the context of the previous search query presumably leading to the cognitive burden of synthesizing disjointed information.

The larger word count and lexical diversity in inputs to LLMs are also expressive indicators that when interacting with the LLM, students are not passively reading information but rather actively processing it by wanting to manipulate it through words such as ``give me an example'' thereby engaging in the crucial co-constructive process of learning with a technology-enabled interface. Compared to the static output of search, this co-constructive approach indicates that students have richer interactions with LLMs as students process the LLM-generated output and verbalize a response with higher lexical richness and semantic value to tailor information for their needs rather than expressing themselves via the less lexically rich input possibilities that search affords. It remains to be discovered if this lexically rich input to LLMs amounts to encouraging a think-out-loud method of learning. Prior studies have shown that thinking out loud while learning can be beneficial to self-regulated learning; however, students do not automatically apply this learning technique and need to be taught how to learn \cite{versteeg2021were}. The affordances of LLMs, if amounting to learning by thinking out loud, suggests a shift where an interactive conversational interface, as a result of its mere affordance, engages students in think-out-loud based learning without deliberate effort from students or teachers. 

From an educators' perspective, the conversational log of an LLM interaction serves as a valuable window into a student's thought process, allowing educators to see how they grapple with new concepts in real time leading to appropriate interventions without invasive technology. Conversely, the search process is a black box. The act of clicking a link reveals little, leaving the educator unable to differentiate between a student who is actively synthesizing information and one who is merely scanning text.

\subsection{Output and outcomes remain surprisingly similar while students may risk a higher change of being caught plagiarizing by using search}

One motivation for students to choose a tool for learning is to translate it into a better grade. When two external school teachers graded the final essays across seven dimensions, we found that the scores across all but the Spelling, Grammar and Mechanics dimension and Flow and Organization differ significantly between pre and post tests across both conditions. The lack of increase in score between pre and post for Flow, Organization, Spelling, Grammar, Mechanics can potentially be explained by the fact that most students were already proficient in the language they were writing. However, did not yet have the topic knowledge leading to a significant difference in Accuracy, Breadth, Depth, Relevancy, and Expertise.

This difference in pre and post scores with both tools  validates of our assumption that students did not know much about the topic before learning it during the experiment and differences in scores in the post-learning phase can be attributed to the tool itself. However, upon analyzing the post-learning essays across the two tools, we only identified a statistically significant difference on the \textit{Accuracy} dimension within our sample with a trending but non statistically significant difference in other dimensions. It was even more surprising that the mean for Accuracy was higher in the LLM condition than the search condition because LLMs have typically been critiqued for hallucinating. Our results also contrast \citet{melumad2025experimental}'s larger-scale study that conducted automated analysis to determine learners demonstrated less depth if they used LLM compared to search. Our results are derived from a different learning setup and evaluation criteria e.g., we used technology as available rather than using constrained data collection platforms, conducted the study in context and in presence of a researcher rather than on online platforms, and employed expert humans to grade output in addition to some automated measures. Crucially, we uniquely connect input characteristics with outcome in this learning context. While we look forward to \citet{melumad2025experimental}'s preprint turn into fully published work to conduct a more thorough comparative analysis, we recognize that this is a growing area of investigation and the learning outcome achieved may largely depend on the background of the sample, grading methods, learning task, and topic of learning among other things.

While human expert and automated analysis using TextEvaluator revealed no difference in the overall characteristics of essays between the two conditions, we found a surprising divergence in detected plagiarism as essays produced after using Search were significantly more likely to be flagged. We attribute this to the nature of the source material: search engines link to existing, indexed content, so even paraphrased student work retains a detectable textual overlap. LLMs, in contrast, generate novel formulations of existing knowledge that do not exist in plagiarism databases.  While empirical research has shown that  neither humans \citep{casal2023can,fleckenstein2024teachers} nor machines \citep{wu2023survey} can reliably detect LLM-generated output, more recent research shows that this finding has not reached the entire student population as some students resist the shift to LLMs for academic work because of the fear of getting caught \citep{divekaraied25}. Ironically, our findings suggests that some current plagiarism tools may inadvertently penalize students for using traditional research methods, while failing to identify students that rely on LLM-generated content. This finding highlights the need to increase AI literacy and also reimagine tools to create more accountability in assessment.

\subsection{LLMs may shift the cognitive load, but do not yield to better outcomes}

\citet{kosmyna2025your}'s preprint discusses using electroencephalography (EEG) to identify brain activity and found that participants writing an essay with LLMs displayed lower brain connectivity than those who used search. Our findings connect it with its manifestation in interaction characteristics and articulating an Interaction-Outcome paradox. We see that when we compared inputs, learners have richer interactions with LLMs as determined by higher lexical richness, number of inputs, and types of inputs. This indicates that rather than spending cognitive resources in   synthesizing complex information like with search, learners are able to interact more richly with LLMs by precisely defining their needs and controlling the output. This potentially shifts their cognitive resources from intrinsic cognitive load to germane cognitive load. For example, instead of synthesizing disjoint information foraged through multiple links via search and looking for a technical explanation that meets them at their level of understanding, learners are able to shift that cognitive load to more explicitly recognizing their own knowledge gap and verbalizing their needs in a prompt and then focus on  reading text that is personalized to them. We note that recognizing one's own knowledge gap, verbalizing it, and finding strategies to bridge it requires and develops metacognitive skills that are useful to learning \cite{efklides2009role}.  

This shift aligns with students creating their own concept of Zone of Proximal Development (ZPD) \citep{vygotsky1978mind} which states that optimal learning happens when the content for the learner is neither so easy that students are disengaged and nor so hard that students are not able to process it. Instead, the difficulty level should lie in the sweet spot where students can build on prior knowledge incrementally with some external support.  An LLM, with its capacity for personalized and adjustable output, theoretically allows a learner to create their ZPD. Therefore, it is possible that learning with LLMs as compared to search, can lead to similar understanding of a topic with less cognitive load, even if reducing brain connectivity per prior research \citep{kosmyna2025your}. However, that freed up cognitive space is potentially being used to find knowledge gaps, verbalize it, and articulate strategies that will lead to bridging the gap as evidenced by the richer and denser input characteristics (e.g., word count, CTTR, type of input request). This potential reorientation of cognitive load, from foraging to articulation of metacognitive awareness and prompting without any visible gains in outcomes, is our best attempt at the explaining the core of the Interaction-Outcome Paradox.  Testing this hypothesis directly presents a compelling avenue for future research, using methods such as think-aloud protocols or cognitive load measures like NASA-TLX or EEGs to further unpack this cognitive shift.


\subsection{Implications for Educational Technology and Interfaces}

 First, we remind ourselves that the most popular tools like Google or ChatGPT are not primarily built to optimize learning. Rather, at the core, Google retrieves information based on page rank and keyword similarity \citep{rogers2002google} whereas LLMs predict the next word based on patterns learned from large training data \citep{vaswani2017attention}. Irrespectively, practitioners and students have retrofitted these tools to fit their learning needs and used it at scale prompting research paradigms like Search As Learning \citep{vakkari2016searching,ghosh2018,vakkari2016searching}. We now shift towards a world where researchers must focus on LLM-enabled Conversational Learning and its impact on the education community. Our field is uniquely prepared for this transformation as the HCI/educational technology research community deeply understands what makes an interface more suitable for education by balancing cognitive load with learning outcomes and must deliberate on how  to create tools that add desirable difficulty elements \citep{bjork2011making} to keep a learner in ZPD while taking away the cognitive load that does not lead to learning. We see a positive trend with large-scale LLM products introducing products specific for learning such as Google LearnLM \citep{google_learnlm_2025} and the study mode for ChatGPT \citep{openai_study_mode_2025}. LearnLM \citep{google_learnlm_2025} provides system instructions that prompt an LLM into being a helpful tutor by incorporating some scaffolding and active learning elements phrases in the prompt context such as ``... ask guiding questions to help your students'', or ``... take incremental steps towards understanding''. Initial analysis shows that the subsequent output of these pre-prompted LLMs of  are better rated by pedagogy experts than results generated by an out-of-the-box LLM \citep{team2024learnlm}. However, future research could discover if students engage in active learning or think-aloud methods with LLMs that engage a student in the learning process primarily as a result of the tool's affordance or prefer to just have the answer through the out-of-the-box LLM, what it does to their cognitive and metacognitive processes, and how it will affect the fast-evolving landscape of moving from search to LLM-based learning.

Our research also showed that students engage in more self-aware articulation where they identify their own knowledge gaps and articulate them explicitly through prompts to get the LLM to generate content at a level they understand. This presents an incredible opportunity for educators to engage their students in more mindful learning as the LLMs expect explicit articulation and engagement with the metacognitive process. However, all students may not be equally adept at identifying  or expressing knowledge gaps leading to frustration or over-estimation of expertise \cite{fernandes2025ai} or over-reliance on LLMs \cite{spatharioti2025effects}. Next-gen educators and educational interfaces may then shift their focus to increasing self-aware articulation of knowledge gaps to help students learn via proactive interfaces. 

Further, we note that students who expressed their knowledge gaps are likely engaging in an otherwise uncomfortable activity as it is often hard to loudly admit to someone of their own shortcomings. This willingness of expression with LLMs presents another opportunity for interaction researchers to create safe learning spaces for intellectual risk taking. However, realizing this potential requires navigating significant challenges. The safe space paradigm is immediately compromised if the LLM provides hallucinatory or unsafe content, eroding the very trust that enables student vulnerability. Further, this personalized safe space opens the possibility of echo chambers and comes at the cost of creating a distance within the educational community by robbing us of engaging in a co-constructive opportunity with other humans \citep{divekaraied25}.






\section{Limitations}
We recognize that identifying the learning and productivity differences with new LLM-based tools is an ongoing topic where findings might depend on the type of task, difficulty, evaluation criteria, sample characteristics, sample size, and other instruments such as automated text and plagiarism analysis tools used in the study. For example, our study recruited N=20 participants from a university with primarily business-oriented majors yet strong familiarity with LLM-based tools such as ChatGPT trying to learn technical topics. Further, while the type of task we designed in this paper reflects a reasonable learning path, it does not reflect all types of learning and education. For example, our study allowed note-taking to reduce the focus memorization and instead explore synthesis and writing. However, memorization and synthesis from memory is an essential part of learning that future research could explore.  Future research will explore repeated longitudinal studies and randomized control trials,  across wider samples and timelines to establish longitudinal effects, and use a broader variety of automated and human text analysis tools. Our paper adds to the body of literature of outcomes while uniquely connecting it with input characteristics, but a definitive answer on learning gains may not be available until future research  repeats the experiment across various conditions and coalesces the various findings into a systematic literature review. Further, we do not explore other factors that affect learning such as intrinsic motivation towards the topic or tool usage characteristics in the wild. While our controlled setting allows us to explore objective learning gains, it is important in future research to also connect it with outcomes driven by hedonic activity usage such as increased engagement because of an easy or gamified user experience  as it can affect tool uptake, eventually leading to better outcomes as a result of engagement alone.

\section{Acknowledgements}
The authors thank Faculty Affairs Committee and Valente Center at Bentley University for funding. We thank Dr. Keelan Evanini and Prof. Roland Hübscher for their feedback and support during this work.
\section{GenAI and Technology Transparency Statement}
All statistical analysis were done using STATA and JASP. GenAI (GPT) was used to assist with Python code (SNS library) used for creating more compact visualizations. All visualizations were cross-checked to be correct by comparing to monochrome JASP output to prevent errors. We used Google Scholar and Consensus to exhaustively search relevant literature.  Authors used Visual Studio Code to write the manuscript in LaTeX and LTeX package to spellcheck writing. Grammarly, GPT, and Claude was used to polish wording in certain sentences. After using this tool/service, the author(s) reviewed and edited the content as needed and take(s) full responsibility for the content of the published article. Raw files and GitHub commit history can be made available upon reasonable request to show incremental writing progress.

\section{Conclusions and Future Work}

This study sheds critical light on the ongoing debate regarding LLMs' impact on learning outcomes compared to search, exploring an Interaction-Outcome Paradox. Through within-subjects experiment with LLM and traditional search interfaces, we found LLM interactions to be considerably richer: involving more words, exhibiting higher lexical diversity, and featuring unique prompt-based requests absent in search-driven queries many indicating that LLM interfaces afford and elicit verbalized self-awareness of the knowledge gap and strategies to bridge the gap. In contrast, search interactions remain characterized by shorter, keyword-based queries. Despite this richer interaction, our findings indicate that LLMs do not broadly yield superior learning outcomes or significantly different output. This disconnect—the Interaction-Outcome Paradox—suggests a fundamental reorientation of cognitive effort. We propose that cognitive demand shifts from synthesizing complex information (via search) to increasing students' articulation of their metacognitive awareness i.e., recognizing their understanding, the gap in the knowledge, and expressing it in a prompt to tailor LLM output. While prior research has linked LLM usage to lower cognitive activity levels, we add nuance by articulating these cognitive shifts and their manifestation in interaction patterns. Crucially, we demonstrate that this shift in cognitive demand does not directly equate to poorer or richer actual learning outcomes. Instead, we hypothesize that this liberated cognitive capacity can be strategically leveraged for desirable difficulties that deepen learning. This presents a critical opportunity for interaction designers to rethink learning design with LLMs, prioritizing environments that foster active engagement and skill development over mere answer generation. 

Adding another layer to this paradox, our research found that essays produced following traditional search were significantly more likely to be flagged for plagiarism by off-the-shelf detection tools. This suggests that the very nature of synthesizing information from existing, indexed web content leaves students vulnerable to plagiarism accusations. Conversely, because LLMs generate novel textual formulations, they produce content that, while not original to the student, is not easily detected by tools designed to find textual overlap. This places students in a situation where the fear of AI detection could ironically lead them to use traditional methods that are more likely to be flagged as plagiarism. This finding presents an urgent call for the HCI and EdTech communities: we must not only focus on increasing AI literacy, but must also fundamentally reimagine assessment and accountability technologies to be resilient and fair in an age where the lines between synthesis, paraphrasing, and generation are irrevocably blurred.

We situate our findings within the context of prior work, noting that our results sometimes align with and sometimes diverge from existing literature. We urge readers and researchers to recognize that outcomes in this field are highly context dependent. The limitations of our study including sample, task, and methodological constraints are discussed in detail in a prior section

\bibliographystyle{elsarticle-harv.bst}
\bibliography{sample-base}


\end{document}